# Uncertainty Quantification-Incorporated Ensemble Model for Personalized Prediction of Severe Radiotherapy-Induced Immunosuppression Among Esophageal Cancer Patients

Baode Gao, MS,[1,2] Yan Chu, PhD,[1] Zongsheng Hu, PhD,[1] Madison E. Grayson, BS,[1] Peter S.N. van Rossum, MD, PhD,[3,4] Clemens Grassberger, PhD,[5,6] Heiko Enderling, PhD,[3,7] Harald Paganetti, PhD,[5,8] Ibrahim Chamseddine, PhD,[5,8] Steven H. Lin, MD, PhD,[3] Brian P. Hobbs, PhD,[9] Radhe Mohan, PhD,[1] Yiqing Chen, PhD[1,10]

1. Department of Radiation Physics, The University of Texas MD Anderson Cancer Center, Houston, Texas
2. Department of Biostatistics and Data Science, University of Texas Health Science Center, Houston, Texas
3. Department of Radiation Oncology, The University of Texas MD Anderson Cancer Center, Houston, Texas
4. Department of Radiation Oncology, Amsterdam UMC, Amsterdam, The Netherlands
5. Department of Radiation Oncology, Massachusetts General Hospital, Boston, MA
6. Department of Radiation Oncology, University of Washington, Seattle, United States
7. Institute for Data Science in Oncology, The University of Texas MD Anderson Cancer Center, Houston, Texas
8. Harvard Medical School, Boston, MA
9. Telperian, Austin, TX
10. Department of Epidemiology and Biostatistics, Texas A&M University, College Station, TX

# Abstract

**Purpose**: Severe radiation-induced lymphopenia (RIL), specifically Grade 4 RIL (G4RIL), is a frequent immune toxicity and a strong predictor of poor survival for esophageal cancer patients undergoing concurrent chemoradiation therapy. This study aimed to develop an *uncertainty-incorporated* ensemble machine learning model to predict Absolute Lymphocyte Count (ALC) nadir, leveraging baseline clinical and dosimetric variables, in order to support treatment planning decisions such as selection for proton therapy or dose distribution optimization to mitigate the risk of severe RIL. Quantifying the uncertainties associated with model predictions is essential for informing clinical decision-making and guiding interventions aimed at improving patient outcomes.

**Methods**: A cohort of 1491 esophageal cancer patients treated with photon or proton therapy between July 2001 and April 2022 was analyzed. A weighted ensemble model was constructed using clinical features and dosimetric variables, including "composite dosimetric score" (CDS) derived from radiation dose distributions for each patient. To quantify individual-level predictive uncertainty, Cross-Residual Conformal Prediction (CRCP) was developed and compared with standard conformal prediction methods.

**Results**: Compared to the photon therapy cohort, the proton therapy group maintained a significantly higher mean ALC nadir of 0.33 K/μL (SD 0.20) compared to 0.25 K/μL (SD 0.16). The ensemble model achieved a mean absolute error of 0.0926 K/μL for ALC nadir prediction and an ROC-AUC of 0.78 for G4RIL classification. Baseline ALC, Planning Target Volume (PTV), and CDS were identified as the most influential predictors. CRCP demonstrated overall strong performance among the methods compared, achieving competitive prediction interval widths with mathematically guaranteed coverage and reflecting a reasonable trade-off between interval width and coverage accuracy.

**Conclusions**: This application of CRCP in radiation oncology maintains high predictive accuracy through ensemble learning while providing prediction intervals for the expected range of ALC nadir, enabling safer and more personalized clinical decision-making.

# Introduction

Esophageal cancer (EC) is an aggressive malignancy for which radiation therapy (RT) is an indispensable component of treatment [1, 2]. However, lymphocytes, which play a critical role in antitumor immunity, are highly radiosensitive and are easily depleted by low and intermediate radiation doses [3, 4]. Consequently, radiation-induced lymphopenia (RIL) is a frequent immune system toxicity for patients undergoing concurrent chemoradiation therapy (chemoRT). Severe RIL, defined as grade 4 RIL (G4RIL) with an absolute lymphocyte count (ALC) of 200 cells per microliter (0.2 K/μL) or less, is a strong predictor of poor survival [5]. Recent analysis has established that G4RIL acts as a causal mediator of overall survival in EC patients [6], and that utilizing highly conformal modalities like proton beam therapy (PBT) significantly mitigates the risk of G4RIL compared to photon-based intensity-modulated radiation therapy (IMRT) [7, 8, 9].

Predictive response modeling based on machine learning (ML) and deep learning (DL) techniques is being investigated widely in radiation oncology [10, 11]. These models are being validated for outcomes predictions in preparation for implementing interventions to optimally mitigate lymphopenia. Examples of such interventions include selection of treatment modality (e.g., photons to protons), optimization of IMRT or intensity-modulated proton therapy (IMPT) plans incorporating immune system as an organ at risk, and reoptimization of plans to adapt to immune system response during the treatment course.

There are many RIL models that leverage both clinical factors and dosimetric features [12]. For example, to address the multicollinearity inherent in highly correlated dose-volume (DV) indices [13] , Chu, et al previously developed the concept of "composite dosimetric score" (CDS) [14]. Using nonnegative matrix factorization, the CDS condenses the dose distributions of relevant immune organs at risk—such as the heart, lungs, spleen and the remainde of the irradiated body volume —into a single, predictive variable for G4RIL classification. In another model, instead of condensing variables, Zhu et al. proposed a two-input channel hybrid deep learning model for G4RIL prediction [15]. This framework utilizes the entire set of DVH parameters by employing a stacked bi-directional long-short term memory (LSTM) neural network to independently process sequential dosimetric information. Simultaneously, a multilayer perceptron structure evaluates non-dosimetric clinical profiles. By concatenating the processed information from both parallel channels, this deep learning model effectively mitigates over-fitting and multicollinearity, demonstrating superior predictive performance over traditional statistical and machine learning approaches. These predictive models may be used in clinical decision-making much like traditional DV constraints are utilized to guide radiation treatment plannin. By predicting the risk of severe RIL either prior to treatment or after the first few fractions, clinicians can make critical, proactive decisions, such as selecting highly conformal RT modalities like PBT for high-risk patients or dynamically reoptimizing dose distributions to better spare immune organs [6, 11, 14]. In the broader context of personalized medicine, these dynamic predictive frameworks lay the conceptual groundwork for digital twin technologies [16, 17], which can simulate and dynamically adapt patient-specific treatment trajectories to proactively manage immune toxicity.

In medical applications, a predicted variable without quantified uncertainty is often insufficient for reliable decision-making. In order for predictions to be useful in the clinic, the rigorous quantification of patient-specific predictive uncertainty is essential. The goal of this project is to develop an uncertainty-incorporated ensemble model that predicts ALC nadir, while concurrently quantifying the degree of uncertainty in these predictions. Our approach utilizes an ensemble model [18] that combines the predictions of multiple individual ML models to mitigate the weaknesses and biases of any single model, yielding a more robust, accurate, and reliable prediction when applied to new patient data. To quantify individual patient-level predictive uncertainty, we utilized conformal prediction (CP) in a novel radiation oncology setting [19]. CP is a distribution-free ML technique that provides distribution-free prediction intervals with finite-sample coverage guarantees under the assumption of exchangeability between calibration and test data.

# Materials and methods

## Study patients

This retrospective study was was conducted based on a protocol approved by the IRB of our institution, the requirement for informed consent was waived in accordance with applicable institutional policies and regulations. We curated data for 1491 patients diagnosed with esophageal cancer (EC) who received concurrent chemoradiotherapy (chemoRT) between July 2001 and April 2022. Relevant inclusion criteria included 1) the availability of baseline Absolute Lymphocyte Count (ALC), and documentation of at least three weekly ALC values during RT; and 2) the availability of Dose-Volume Histogram (DVH) data. After excluding patients with missing baseline ALC and less than 3 weekly ALC values, the final cohort of 1395 patients included 956 patients treated with photon and 439 patients treated with proton radiation. The outcome of interest for studies utilizing this cohort was the ALC nadir during RT and Grade 4 Radiation-Induced Lymphopenia (G4RIL).

## Univariate Statistical Analysis

To compare demographic and clinical characteristics between the proton and photon groups, T-tests were employed for continuous data (mean, SD), while chi-squared tests were applied for categorical variables (frequencies, percentages). All comparisons were two-sided, and considered a $p$-value less than $0.05$ as evidence of a significant difference.

## ALC nadir prediction and G4RIL classification

The demographic characteristics included body mass index (BMI), age, gender and clinical variables included ECOG, smoking, T stage, N stage, clinical stage, Planning Target Volume (PTV), and baseline ALC; whereas the dosimetric indices included whole body, lung, spleen, heart DVHs. They were all considered in modelling. To solve the multicollinearity issue in dosimetric data, "Composite Dosimetric Score" (CDS) [12] was employed. The CDS is computed utilizing Nonnegative Matrix Factorization (NMF), which is an unsupervised machine learning technique. Dose-volume indices of the lung, heart, and spleen were used to compute CDS values for each of these structures, and combined into a single CDS value per patient, effectively reducing multicollinearity. All patients with missing body DVHs were removed and the 966 patients remaining for used for predictive modeling. Outcome ALC nadir count was log-transformed. Both ALC nadir prediction and G4RIL classification were modeled by weighted ensemble model with architecture in Figure 1.

The evaluation metrics for ALC nadir prediction are mean absolute error and root mean squared error. For G4RIL classification, AUC and F1-score were computed and applied.

## Uncertainty quantification

In this study, conformal inference based uncertainty quantification was applied specifically to the continuous ALC nadir prediction. To quantify individual-level predictive uncertainty, we developed Cross-Residual Conformal Prediction (CRCP) with weighted ensemble model (Figure 2). CRCP is a distribution-free method that combines residual-scaled conformal prediction with $K$-fold cross-calibration to improve data efficiency and stability.

Let the dataset be partitioned into $K = 5$ mutually exclusive folds $\{\mathcal{I}_1, \dots, \mathcal{I}_5\}$. For each fold $k$, the complement set $\mathcal{I}_{-k} = \bigcup_{j \neq k} \mathcal{I}_j$ is used to train (i) an ensemble prediction model $\hat{f}^{(-k)}(\cdot)$ and (ii) a residual model $\hat{r}^{(-k)}(\cdot)$, while fold $\mathcal{I}_k$ serves as the calibration set. For each observation $i \in \mathcal{I}_k$, the scaled nonconformity score is computed as

$$s_i^{(k)} = \frac{\left| y_i - \hat{f}^{(-k)}(x_i) \right|}{\hat{r}^{(-k)}(x_i)}.$$

The pooled calibration set is obtained by

$$S = \bigcup_{k=1}^{5} \left\{ s_i^{(k)} : i \in \mathcal{I}_k \right\},$$

The conformal threshold is defined as

$$q = \text{ quantile } \left( S, \frac{(1-\alpha)(n+1)}{n} \right),$$

where $\alpha$ denotes the nominal error rate. For a new individual with covariates $x_{\text{test}}$, the CRCP prediction interval is obtained by:

$$PI(x_{\text{test}}) = \left[\hat{f}(x_{\text{test}}) - q\hat{r}(x_{\text{test}}), \hat{f}(x_{\text{test}}) + q\hat{r}(x_{\text{test}})\right],$$

where $\hat{f}(\cdot)$ and $\hat{r}(\cdot)$ are ensemble prediction model and ensemble residual model trained on the combined set of the training set and the calibration set for inference. Empirical coverage of CRCP intervals was assessed on the independent test set.

The primary goal of uncertainty quantification using conformal prediction is to achieve a coverage rate close to the target confidence level while maintaining a short average length as a measure of how well uncertainty estimates are calibrated. The dataset was split into training-calibration set and test set with ratio 80:20. 5-Fold Cross-Calibration was employed to this training-calibration set. The test set was only used for the final performance evaluation. Three conformal prediction methods (conformal prediction, residual conformal prediction, and CRCP) performance were compared.

# Results

## Univariate Statistical analysis

Table 1 shows descriptive statistics for the cohort.The study analyzed 1395 patients, comparing photon therapy ($n = 956$) with proton therapy ($n = 439$). The initial comparison showed that the mean value of ALC nadir in the proton group (0.33) was significantly higher than that in the photon group (0.25; p<0.001) even though the baseline ALC were comparable. Proton patients were modestly older (66.42 years comparing to 62.14 years for photon patients, P<0.001). As shown in table 1, PTVs, CTVs and GTVs, on average were smaller for proton patients.

## ALC nadir prediction and G4RIL classification

For ALC nadir prediction, the weighted ensemble model was constructed from a diverse set of base learners including NeuralNetTorch, NeuralNetFastAI, LightGBM, CatBoost, ExtraTrees, RandomForest and XGBoost. After greedy forward selection, the final weighted ensemble model is composed of CatBoost (29.2%), NeuralNetTorch (45.8%) and NeuralNetFastAI (25.0%). In the independent test set, the weighted ensemble achieved a mean absolute error of $0.0926\ K/\mu L$ and a root mean square error of $0.1328\ K/\mu L$. It performed better than all the individual base models (shown in supplement). The SHAP value was employed for model explanation. As shown in Figure 3(a), the three most important features were baseline ALC, PTV, and body CDS-. Body CDS- represents CDS calculated using the body DVH excluding GTV, spleen, heart and lung DVH. Body CDS+ represents CDS calculated using the body DVH excluding GTV only. These features have the widest spread of SHAP values (Figure 3(b)), indicating their substantial and heterogeneous effects on individual-level prediction. Higher baseline ALC values are consistently associated with positive SHAP values, strongly increasing the predicted ALC nadir, which indicates better lymphocyte preservation. Larger PTVs are associated with predominantly negative SHAP values, reflecting a substantial decrease in the predicted ALC nadir as irradiated volumes increase. Lower CDS values are associated with positive SHAP contributions and higher predicted ALC nadir. After combining dose distributions from different organs to one sigle CDS, the body CDS+ SHAP indicates that it has greater predictive importance than PTV (Figure 3(c)).

In G4RIL classification, NeuralNetTorch, NeuralNetFastAI, LightGBM, CatBoost, ExtraTrees, RandomForest, and XGBoost, WeightedEnsemble were included as base learner for the weighted ensemble model. CatBoost (16.7%), XGBoost (11.1%), NeuralNetTorch (44.4%) and NeuralNetFastAI (27.8%) were finally selected. The model achieved F1-score 0.639 and ROCAUC 0.78. The influence of the features in G4RIL classification (Figure 4) are the reverse of those in ALC nadir prediction. This is because the lower ALC nadir implies the higher risk of G4RIL. The most significant predictors of G4RIL are the baseline ALC, body CDS- and PTV. A higher baseline ALC is the primary driver of lower G4RIL risk, whereas a large PTV is associated with higher risk of G4RIL. Furthermore, the lower body CDS- value appears to offer a protection against G4RIL. Notably, after incorporating CDS into the model, body CDS+ emerged as the most important predictor based on SHAP values, highlighting the significance of “dose bath” in inducing lymphopenia.

## Uncertainty quantification

To minimize the average interval length, the empirically calibrated $\alpha$ is utilized during the evaluation phase. Specifically, a smoothing spline is fitted to model the relationship between the nominal $\alpha$ and the empirical coverage rate obtained from the calibration set.

The fitted spline is then used to identify the α value corresponding to the desired confidence level. Table2 summarizes the performance of three conformal prediction methods (CP, RCP, and CRCP) in quantifying uncertainty across three nominal alpha (0.1, 0.05, and 0.01). CRCP provides the tightest prediction intervals across different α values, making it the most efficient method for uncertainty quantification.

The coverage analysis visualization in the supplementary materials illustrates how the constructed prediction intervals capture the actual ALC nadir for individual patients in the testing set at 90% prediction level. Overall, 92.76% of the observed ALC nadir values fall within the corresponding prediction intervals, indicating promising empirical coverage. The width of the prediction intervals varies systematically across the range of ALC nadir values. For patients with low ALC nadir values, the prediction intervals tend to be narrower, reflecting lower predictive uncertainty. Patients with high ALC nadir values exhibit substantially wider intervals. This suggests an increased uncertainty or greater data variability for patients with higher ALC nadir values. The plot in supplementary materials (S2) shows that CRCP is an effective method for providing calibrated uncertainty estimates (92.76% coverage rate in 90% confidence level). The method's failures are concentrated in the region of the highest ALC nadir values, arguably due to higher ALC nadir corresponding to lower risks of severe lymphopenia. These failure cases are of comparatively low clinical consequence. The model's stronger performance in the low-nadir range is directly relevant to identifying patients at risk for G4RIL who may benefit from proactive intervention. Conversely, the model demonstrates robust predictive interval in the lower ranges of the ALC nadir. This performance characteristic is highly advantageous from a clinical perspective, as low ALC nadir values indicates severe lymphopenia and accurate predictions are most critical for guiding interventions in such cases.

# Discussion

In this study, we developed and validated an uncertainty-incorporated ensemble machine learning model to predict ALC nadir in esophageal cancer patients. A significant finding of the univariate analysis was that proton therapy patients exhibited a significantly higher mean ALC nadir compared to photon patients (0.25 K/μL (SD 0.16)), despite similar baseline ALC levels. SHAP value analysis identified baseline ALC, Planning Target Volume, and Composite Dosimetric Score (CDS) as the most influential predictors for prediction of severe RIL. Notably, unlike baseline ALC or PTV, CDS is derived directly from the planned dose distribution and is therefore the only major predictor that can be actively modified during treatment planning, offering a concrete, actionable lever for reducing the risk of severe RIL. To assess whether these modality-specific differences affected model evaluation, we conducted subgroup analyses on the test cohort. Subgroup analysis revealed notable performance variations between treatment modalities, particularly in G4RIL risk classification. While the model maintained reasonable performance in ALC nadir prediction across both cohorts (proton MAE: 0.1139 vs. photon MAE: 0.0835), its classification capability for G4RIL was distinctly higher in the photon-only cohort (ROC-AUC: 0.809, F1-score: 0.705) compared to the proton-only cohort (ROC-AUC: 0.589, F1-score: 0.333). This discrepancy reflects baseline distribution shifts and sample size differences to proton therapy.

The primary novelty of this work lies in the rigorous application of conformal prediction for uncertainty quantification. We compared standard Conformal Prediction, RCP, and CRCP. Analyses showed that CRCP offered the most efficient uncertainty quantification, producing the tightest prediction intervals while maintaining valid empirical coverage probability. Of interest, prediction intervals are generally narrower for patients with lower predicted ALC nadir values, indicating lower predictive uncertainty, whereas they become progressively wider for patients with higher predicted ALC nadir values, reflecting greater uncertainty. Because lower ALC nadirs are associated with an increased risk of severe (G4) lymphopenia, this finding has important clinical implications. Specifically, predictions identifying patients at the highest risk of severe lymphopenia can be interpreted with greater confidence, thereby supporting more informed decisions regarding dosimetric interventions. Such interventions may include reoptimization of the treatment plan or selection of IMPT instead of IMRT or VMAT when appropriate to reduce the risk of treatment-induced lymphopenia.

It is important to note that, under current clinical practice, uncertainties associated with dosimetric parameters used for treatment planning, such as dose-volume constraints and mean, minimum and maximum doses (which are implicitly model predictions), are generally not explicitly incorporated into treatment planning and optimization, treatment plan evaluation, or decisions regarding adaptive or dosimetric interventions. In this regard, the presented uncertainty quantification framework provides an important opportunity to advance the state of the art in radiotherapy by enabling more informed, risk-aware, and personalized clinical decision-making.

There are limitations to this study. First, it is retrospective and sourced from a single institution, which may introduce selection bias. Second, the modeling cohort was reduced due to missing body DVH data, as the model cannot make predictions for patients lacking this information. It is possible that these data are not missing at random, which could introduce further selection bias. Another limitation is the consistency of the ALC nadir measurement. Since the frequency of blood tests varied among patients, the true nadir might have been missed in patients with fewer measurements compared to those with more frequent testing. This inconsistency could introduce noise into the ground truth labels used for model training. Finally, although the CRCP method provided the uncertainty quantification, the data points outside the interval were concentrated in the high-ALC nadir region, suggesting that the model may struggle to capture the variability of patients who maintain exceptionally high lymphocyte during treatment.

## Conclusions

To our knowledge, this is among the first applications of uncertainty quantification in radiation oncology to provide personalized, mathematically guaranteed prediction uncertainty intervals for RIL. Our results demonstrate that integrating conformal prediction with an ensemble learning framework not only yields highly accurate predictions of ALC nadir and the risk of severe lymphopenia, but also quantifies the uncertainty associated with each individual prediction. This additional information provides an objective measure of confidence that can facilitate risk stratification and support clinical decision-making regarding interventions to mitigate RIL, including treatment plan reoptimization and selection of proton therapy when appropriate.

This work represents an important step toward raising the awareness of uncertainty in radiotherapy, especially with regard to dosimetric parameters. Under current clinical practice, uncertainties associated with predictive models and the dosimetric parameters used to guide treatment planning and intervention decisions are generally not incorporated explicitly into the clinical decision-making process. By providing individualized uncertainty estimates in addition to point predictions, our framework enables more informed, risk-aware, and personalized treatment decisions while improving the transparency and interpretability of predictive models. Beyond RIL prediction, the proposed methodology is broadly applicable to other predictive models in radiation oncology and has the potential to facilitate the development of other clinically-deployable artificial intelligence–based decision support systems.

# Figure captions

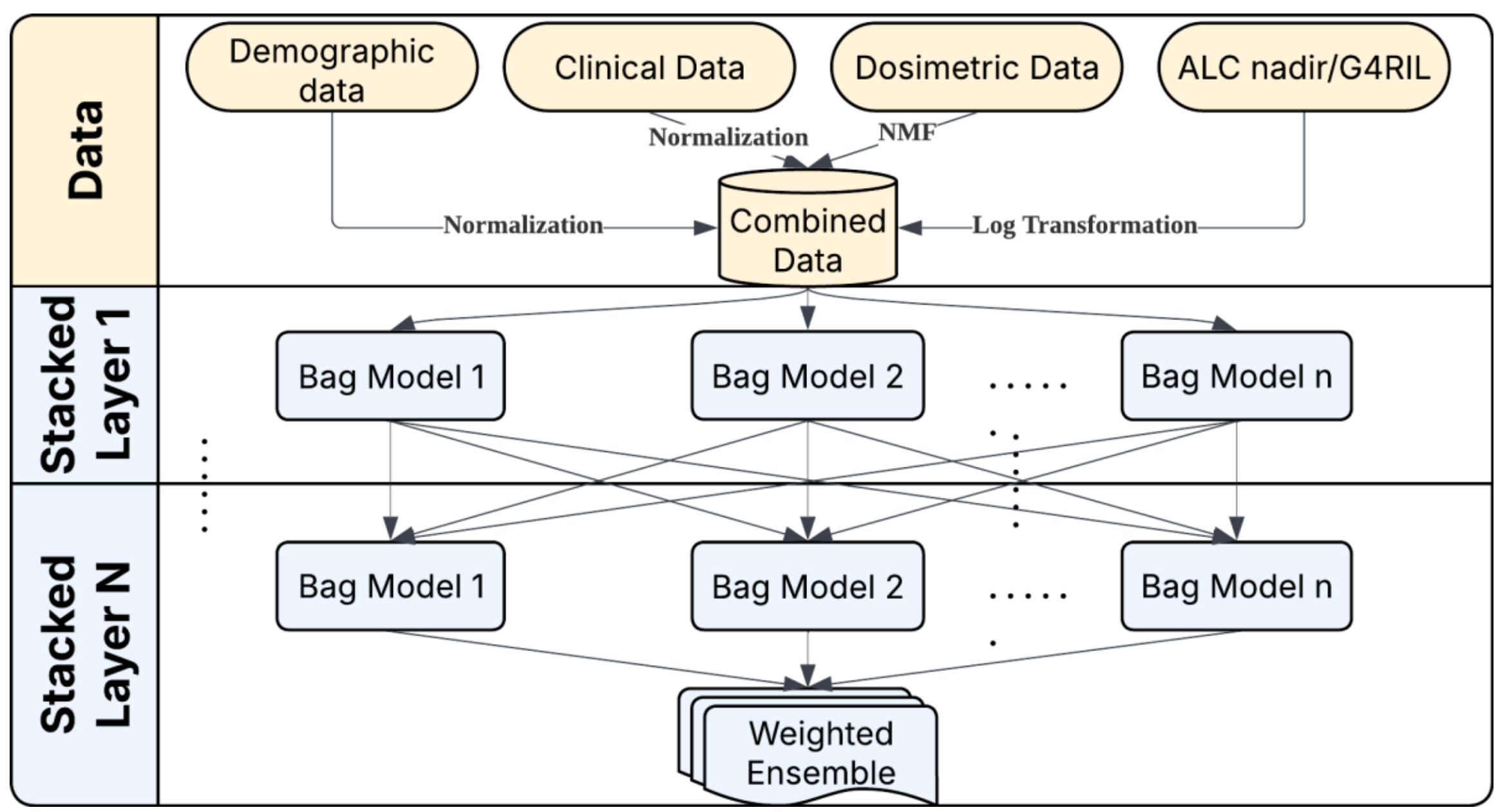

*Figure 1 Architecture of ensemble model. Greedy forward selection is used to construct the weighted ensemble model. It first selects the model with the highest score in the validation set. Then, it attempts to incorporate each of the other models one by on to see which one can maximize the overall validation set score. Only stacked layer 1 takes patients' feature as input. Other stacked layers take the predictions made by the previous layer and uses those predictions as its new input features to learn how to combine the strengths and weaknesses of the base models. 5-fold cross-validation was employed for better data utilization and reducing overfitting.*

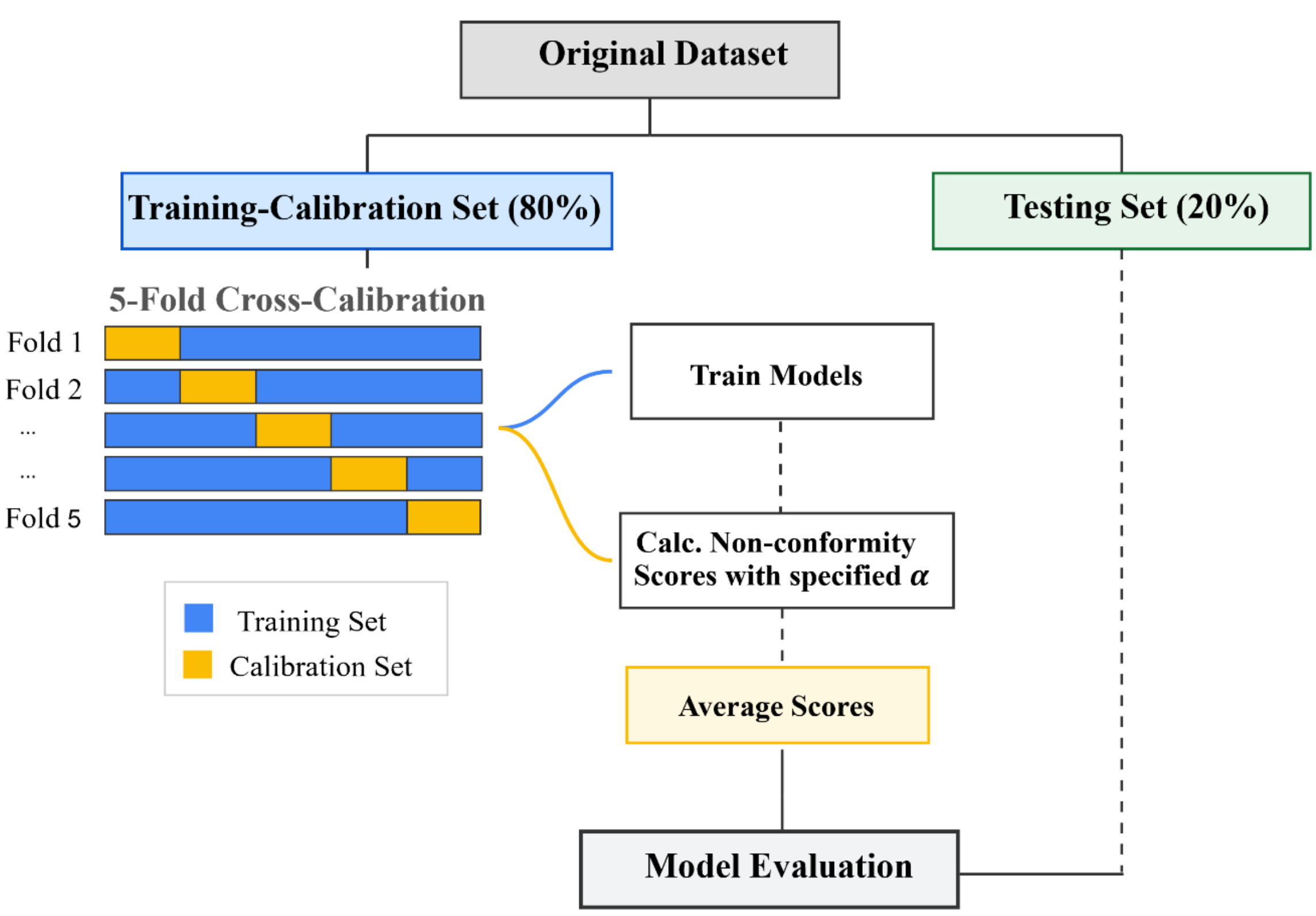

*Figure 2 Workflow for Conformal Prediction*

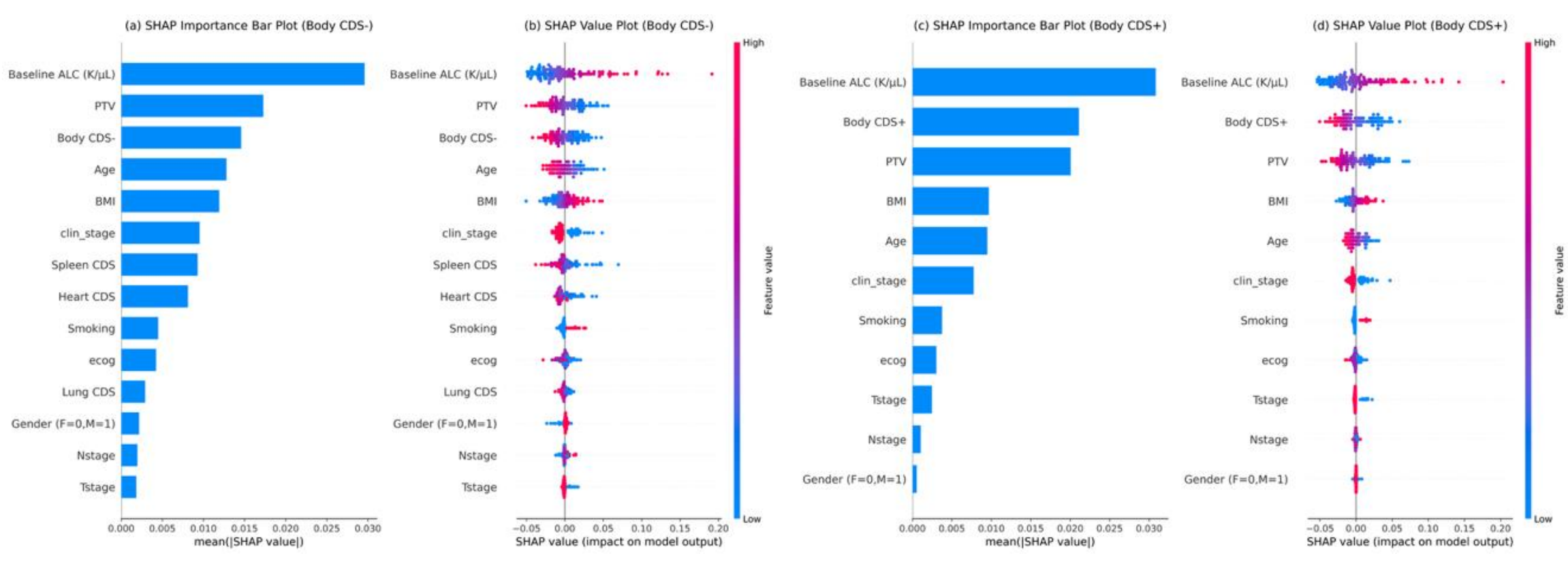

*Figure 3 Feature importance and SHAP values for ensemble model in ALC nadir prediction. Body CDS+ represents body DVH excluding GTV. Body CDS- represents body DVH excluding GTV, spleen, heart and lung DVH.*

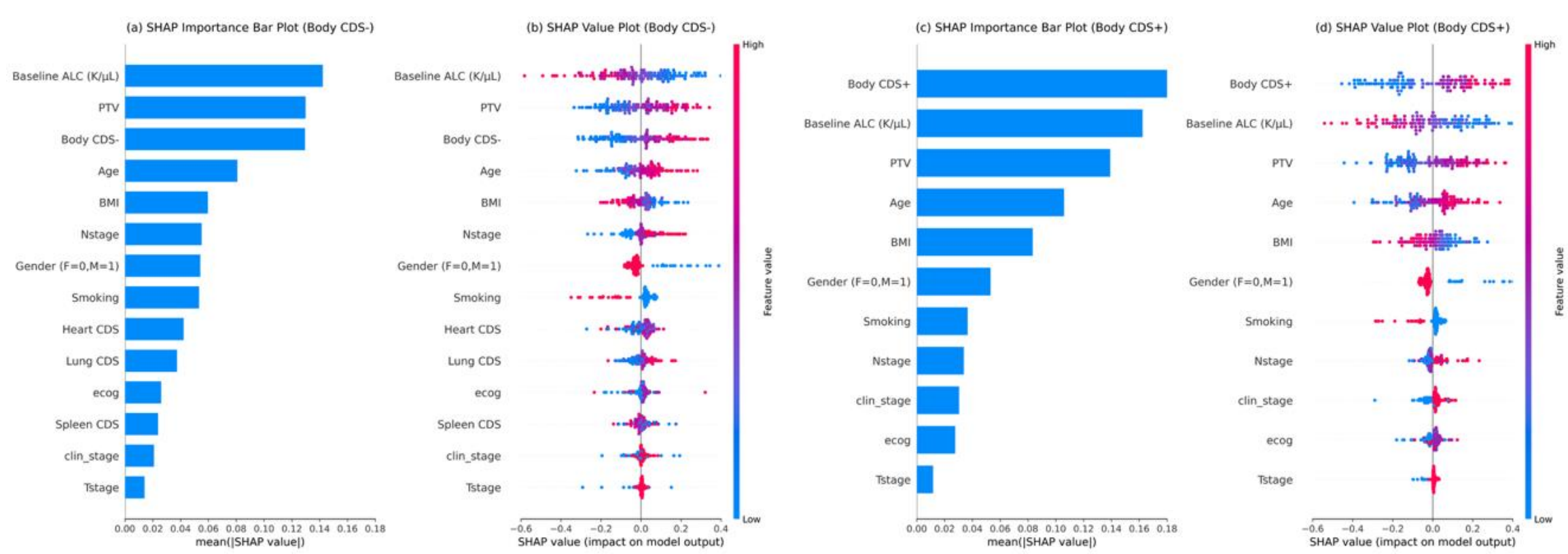

*Figure 4 Feature importance and SHAP values for ensemble model in G4RIL classification. Body CDS+ represents body DVH excluding GTV. Body CDS- represents body DVH excluding GTV, spleen, heart and lung DVH.*

| Variable / Level | Overall | Photon | Proton | P-value |
|---|---|---|---|---|
| | **(n=1395)** | **(n=956)** | **(n=439)** | |
| Age, years mean (SD) | 63.49 (10.67) | 62.14 (10.72) | 66.42 (9.97) | <0.001 |
| BMI, $KG/M^2$ mean (SD) | 26.49 (5.94) | 26.40 (6.21) | 26.62 (5.23) | 0.515 |
| **Sex, n (%)** | | | | 0.233 |
| Female | 226 (16.2) | 163 (17.1) | 63 (14.4) | |
| Male | 1172 (83.8) | 793 (82.9) | 376 (85.6) | |
| **ECOG Status, n (%)** | | | | 0.028 |
| 0 | 458 (32.8) | 306 (32.0) | 151 (34.4) | |
| 1 | 847 (60.6) | 574 (60.0) | 271 (61.7) | |
| 2 | 91 (6.5) | 75 (7.8) | 16 (3.6) | |
| 3 | 2 (0.1) | 1 (0.1) | 1 (0.2) | |
| **Smoking Status, n (%)** | | | | 0.047 |

| Never | 355 (25.4) | 234 (24.5) | 121 (27.6) | |
|---|---|---|---|---|
| Past/Quit | 659 (47.1) | 439 (45.9) | 217 (49.4) | |
| Current, < 1 Pack/Day | 77 (5.5) | 61 (6.4) | 16 (3.6) | |
| Current, ≥ 1 Pack/Day | 102 (7.3) | 82 (8.6) | 20 (4.6) | |
| Smoke (not detailed) | 11 (0.8) | 8 (0.8) | 3 (0.7) | |
| Unknown | 3 (0.2) | 2 (0.2) | 1 (0.2) | |
| **Histology, n (%)** | | | | 0.652 |
| Adeno | 1117 (79.9) | 758 (79.3) | 356 (81.1) | |
| SCC | 270 (19.3) | 191 (20.0) | 79 (18.0) | |
| Other | 11 (0.8) | 7 (0.7) | 4 (0.9) | |
| PTV, $cm^3$ mean (SD) | 654.31 (362.35) | 712.94 (394.57) | 529.38 (237.97) | <0.001 |
| CTV, $cm^3$ mean (SD) | 373.92 (242.844) | 402.68 (267.17) | 316.55 (171.47) | <0.001 |
| GTV, $cm^3$ mean (SD) | 67.16 (58.43) | 70.39 (61.71) | 60.64 (50.62) | 0.011 |
| **Surgery, n (%)** | | | | 0.206 |
| No | 755 (54.0) | 501 (52.4) | 251 (57.2) | |
| Yes | 641 (45.9) | 454 (47.5) | 187 (42.6) | |
| Baseline ALC, $K/\mu L$ mean (SD) | 1.57 (0.60) | 1.59 (0.61) | 1.54 (0.58) | 0.180 |
| ALC nadir, $K/\mu L$ mean (SD) | 0.27 (0.18) | 0.25 (0.16) | 0.33 (0.20) | <0.001 |

*Table 1Characteristics Stratified by RT Group (Photon vs. Proton)*

| | **CP** | **RCP** | **CRCP** |
|---|---|---|---|
| **$\alpha = 0.1$ (90% interval)** | | | |
| Avg. Length | 0.516 | 0.488 | **0.462** |
| **$\alpha = 0.05$ (95% interval)** | | | |
| Avg. Length | 0.631 | **0.532** | 0.541 |
| **$\alpha = 0.01$ (99% interval)** | | | |
| Avg. Length | 0.886 | 0.872 | **0.852** |

• **CP**: conformal prediction

• **RCP**: residual conformal prediction

• **CRCP**: cross residual conformal prediction

*Table 2 Summary of Conformal Prediction Evaluation Results, $K/\mu L$*